\documentclass[a4paper]{jpconf}
\usepackage{graphicx}
\begin{document}
\title{NEMO~3 double beta decay experiment: latest results}

\author{Alexander Barabash (on behalf of the NEMO Collaboration)}

\address{Institute of Theoretical and Experimental Physics, B. Cheremushkinskaya 25, 117218 Moscow, Russia}

\ead{barabash@itep.ru}

\begin{abstract}
The double beta decay experiment NEMO~3 has been taking data since February 2003. The aim of this 
experiment is to search for neutrinoless decay and investigate two neutrino double beta decay in seven different 
enriched isotopes ($^{100}$Mo, $^{82}$Se, $^{48}$Ca, $^{96}$Zr, $^{116}$Cd, $^{130}$Te and $^{150}$Nd).
After analysis of the data corresponding to 693 days, no evidence for $0\nu\beta\beta$ decay in $^{100}$Mo and 
$^{82}$Se was found. The half-life limits at 90$\%$ C.L. are $5.8\cdot 10^{23}$ y and $2.1\cdot 10^{23}$ y, 
respectively. Using $\sim$ 940 days of data, preliminary results were obtained for $^{48}$Ca, $^{96}$Zr and $^{150}$Nd.
For two neutrino decay half-life values were measured, 
$T_{1/2}(^{48}{\rm Ca}) = [4.4^{+0.5}_{-0.4}(stat) \pm 0.4(syst)] \cdot 10^{19}$ y,
$T_{1/2}(^{96}{\rm Zr}) = [2.3 \pm 0.2(stat) \pm 0.3(syst)] \cdot 10^{19}$ y and 
$T_{1/2}(^{150}{\rm Nd}) = [9.2^{+0.25}_{-0.22}(stat) \pm 0.62(syst)] \cdot 10^{18}$ y. 
Also for $0\nu\beta\beta$ decay the following limits at 90$\%$ C.L. were obtained, 
$> 1.3 \cdot 10^{22}$ y for $^{48}$Ca, $> 8.6 \cdot 10^{21}$ y for $^{96}$Zr and 
$> 1.8 \cdot 10^{22}$ y for $^{150}$Nd.   

\end{abstract}

\section{Introduction}

Interest in neutrinoless double-beta decay has seen a significant renewal in 
recent years after evidence for neutrino oscillations was obtained from the 
results of atmospheric, solar, reactor and accelerator  neutrino 
experiments (see, for example, the discussions in \cite{VAL06,BIL06,MOH06}). 
These results are impressive proof that neutrinos have a non-zero mass. However,
the experiments studying neutrino oscillations are not sensitive to the nature
of the neutrino mass (Dirac or Majorana) and provide no information on the 
absolute scale of the neutrino masses, since such experiments are sensitive 
only to the difference of the masses, $\Delta m^2$. The detection and study 
of $0\nu\beta\beta$ decay may clarify the following problems of neutrino 
physics (see discussions in \cite{PAS03,MOH05,PAS06}):
 (i) lepton number non-conservation, (ii) neutrino nature: 
whether the neutrino is a Dirac or a Majorana particle, (iii) absolute neutrino
 mass scale (a measurement or a limit on $m_1$), (iv) the type of neutrino 
mass hierarchy (normal, inverted, or quasidegenerate), (v) CP violation in 
the lepton sector (measurement of the Majorana CP-violating phases).

The currently running NEMO~3 experiment is devoted to the search for $0\nu\beta\beta$ decay 
and to the accurate measurement of two neutrino double beta decay ($2\nu\beta\beta$ decay) by 
means of the direct detection of the two electrons. This tracking experiment, in contrast to experiments
with $^{76}$Ge, detects not only the total energy deposition,
but other parameters of the process. These include the energy of
the individual electrons, angle between them,
and the coordinates of the event in the source plane. Since June of 2002, the NEMO~3 detector
has operated in the Fr\'ejus Underground Laboratory (France)
located at a depth of 4800 m w.e. Since February 2003, after the final tuning of the 
experimental set-up, NEMO~3 has been taking data devoted to double beta decay studies.
The first obtained results with $^{100}$Mo and $^{82}$Se were published in \cite{ARN04,ARN05a,ARN06,ARN07}.

\section{The NEMO~3 detector}

\begin{figure}
  \begin{center} 
    \includegraphics[height=10cm]{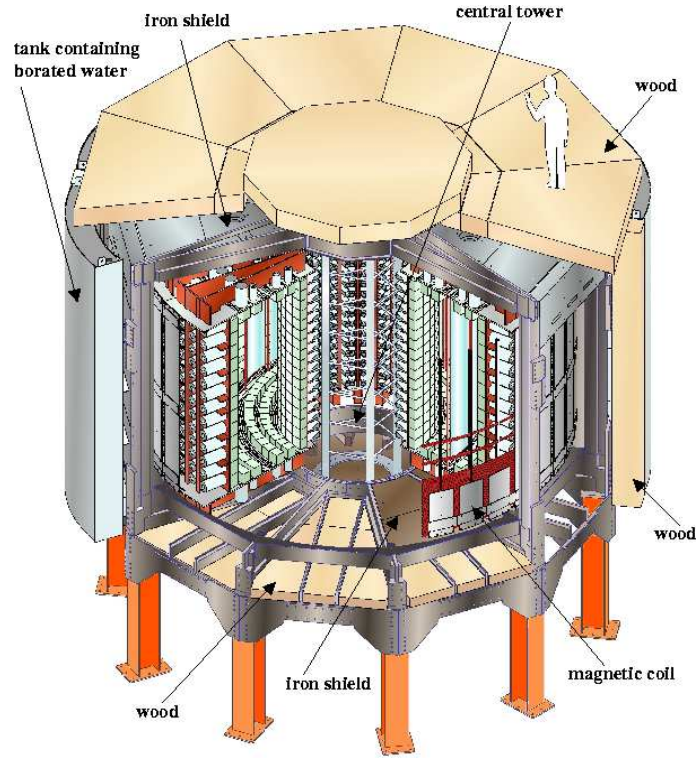} 
  \end{center}
  \caption{A schematic view of the NEMO~3 detector.} 
  \label{fig:nemo3-detector}  
\end{figure} 

The NEMO~3 detector  has three main components, a foil consisting
of different sources of double beta decay isotopes and copper, 
a tracker
made of Geiger wire cells and a calorimeter made of scintillator
blocks with PMT readout, surrounded by a solenoidal coil. The
detector has the ability to discriminate between events of different
types by positive identification of charged tracks and photons.  
A schematic view of the NEMO~3 detector is
shown in Fig.~\ref{fig:nemo3-detector}. 

The  NEMO~3 detector is cylindrical in  design and is composed of
twenty equal sectors.  The external dimensions of the detector with
shields are about 6 m in diameter and 4 m in height.  
NEMO~3 is based on the techniques tested on previous incarnations of
the experiment NEMO~1 \cite{DAS91}, and NEMO~2 \cite{ARN95}. 

The  wire chamber is  made of  6180 open  octagonal drift  cells which
operate in Geiger  mode (Geiger cells).  A gas  mixture of $\sim$ 95\%
helium, 4\% ethyl-alcohol, 1\% argon and 0.15\% water at 10 mbar above
atmospheric pressure is  used as the filling gas  of the wire chamber.
Each  drift  cell provides  a  three-dimensional  measurement of  the
charged particle tracks by recording the drift time and the two plasma
propagation  times.  The  transverse position  is determined  from the
drift  time,  while the  longitudinal  position  is  deduced from  the
difference between  the plasma propagation  times at both ends  of the
cathode wires.   The average vertex  position resolution for  the two-electron
events is $\sigma_t = 0.5$ cm  in the transverse plane of the detector
and $\sigma_l = 0.8$ cm in the longitudinal plane.
The Geiger counters information is treated by the track reconstruction
program   based  on  the   cellular  automaton   algorithm,  described
in \cite{KIS97}. 

The calorimeter, which surrounds the wire chamber, is composed of 1940
plastic   scintillator  blocks   coupled  by   light-guides   to  very
low-radioactivity   photomultiplier    tubes   (PMTs)   developed   by
Hammamatsu.   The energy  resolution FWHM  of  the calorimeter
ranges  from 14.1 to 17.6\% for  1 MeV  electrons, while  the  time
resolution is of 250 ps at 1 MeV.

\begin{table}
\caption{Investigated isotopes with NEMO~3 \cite{ARN05}.}
\begin{center}
\begin{tabular}{cccccccc}
\hline
Isotope & $^{100}$Mo & $^{82}$Se & $^{130}$Te & $^{116}$Cd &
$^{150}$Nd & $^{96}$Zr & $^{48}$Ca  \\
\hline
Enrichment, \% & 97 & 97 & 89 & 93 & 91 & 57 & 73 \\
Mass of isotope, g & 6914 & 932 & 454 & 405 & 36.6 & 9.4 & 7.0 \\
\hline
\end{tabular}
\end{center}
\end{table}

The apparatus accommodates almost 10 kg of different double beta decay
isotopes (see table~1).
Most of  these  isotopes are highly enriched and are
shaped in the  form  of thin  metallic  or composite  foils with  a
density of 30-60 mg/cm$^2$. 
Three sectors  are also
used for external background  measurement and are equipped respectively
with pure Cu (one sector, 621 g) and natural Te (1.7 sectors, 614 g
of $^{nat}$TeO$_2$).  
Some of the sources, including $^{100}$Mo,
have been purified in
order  to  reduce their  content  of  $^{208}$Tl (from $^{232}$Th
and $^{228}$Th, and from $^{228}$Ra with a half-life of 5.75 y) and
$^{214}$Bi (from $^{226}$Ra with a half-life of 1600 y)
either  by  a  chemical  procedure \cite{ARN01},  or  by  a  physical
procedure \cite{ARN05}.  The foils  are placed inside the wire chamber
in the  central vertical  plane of each  sector.  The majority  of the
detector, 12 sectors, is mounted with 6.9 kg of $^{100}$Mo.

The detector is surrounded  by a solenoidal coil which generates
a vertical magnetic  field of 25 Gauss inside the  wire chamber.  This
magnetic field  allows electron-positron identification by measuring
the curvature of their tracks.  The ambiguity of the e$^+$/e$^-$ recognition
based on the curvature reconstruction is 3\% at 1 MeV.
 
The  whole detector  is  covered  by two  types  of shielding  against
external  $\gamma$-rays and  neutrons.  The  inner shield  is  made of
20 cm thick  low radioactivity iron which stops $\gamma$-rays  and slow
neutrons.  The  outer shield is comprised  of tanks filled  with borated
water  on the  vertical walls  and  wood on  the top  and bottom 
designed to thermalize and capture neutrons.

At the beginning of the experiment, the radon inside the tracking chamber, 
and more precisely the its decay product $^{214}$Bi present in its radioactive 
chain was found to be the predominant background. Radon is present 
in the air of the laboratory and originates from the rock surrounding. It 
can penetrate the detector through small leaks. A tent coupled 
to a radon-free air factory was installed around the detector in October 
2004 in order to decrease the presence of radon inside the tracker.

Since February 2003, after the final tuning of the
experimental set-up, NEMO~3 has  routinely been taking data devoted to
double beta  decay studies.  The calibration  with radioactive sources
is carried  out every  6 weeks.  The  stability of the  calorimeter is
checked daily with a laser based calibration system \cite{ARN05}.

The advantage of the NEMO~3 detector rests in its capability to
identify the two electrons from $\beta\beta$ decay and the
de-excitation photons from the  excited state of the daughter nucleus.
The  NEMO~3  calorimeter  also  measures  the detection  time  of  the
particles. The  use of appropriate time-of-flight (TOF) cuts,  in
addition to energy cuts, allows an efficient reduction of all
backgrounds.

A full description of the detector and its characteristics can be found in \cite{ARN05}.

\section{Experimental results} 

A candidate for a $\beta\beta$ decay is a two-electron event which is defined 
with the following criteria: two tracks coming from the same vertex in a source foils, 
the curvature of the tracks corresponds to a negative charge, each track has 
to be associated with a trigged scintillator, and the time-of-flight has to 
correspond to the case of two electrons emitted at the same time from the 
same vertex. The energy deposited in the counter is required to exceed 200 keV, 
except with $^{82}$Se where the threshold was selected to be 300 keV.
In order to suppress backgrounds from $^{214}$Bi decay inside the 
tracking detector, that is followed by $^{214}$Po $\alpha$-decay, it is required that
there is no delayed Geiger cell hit close to the event vertex.

A complete study of backgrounds has been performed to date. The level of each background 
has been directly measured from the data. The first 
run period, from February 2003 to September 2004, with high level of radon is Phase I.  
The radon level inside NEMO~3 during the second run after the installation 
of the radon trapping facility, Phase II (since November 2004 up to now), has been reduced 
by a factor six.

\subsection{Results for $^{100}$Mo ($Q_{2\beta}$ = 3.034 MeV)}

The $2\nu\beta\beta$ decay of $^{100}$Mo has been measured with high accuracy in NEMO~3. 
Fig. 2a  displays the
spectrum of $2\nu\beta\beta$ events
for $^{100}$Mo that were collected over 389 days (Phase I)
\cite{ARN05a}. The angular distribution
(Fig. 2b) and single electron spectrum (Fig. 2c) are also shown.

\begin{figure}
\begin{center}
\includegraphics[scale=0.35]{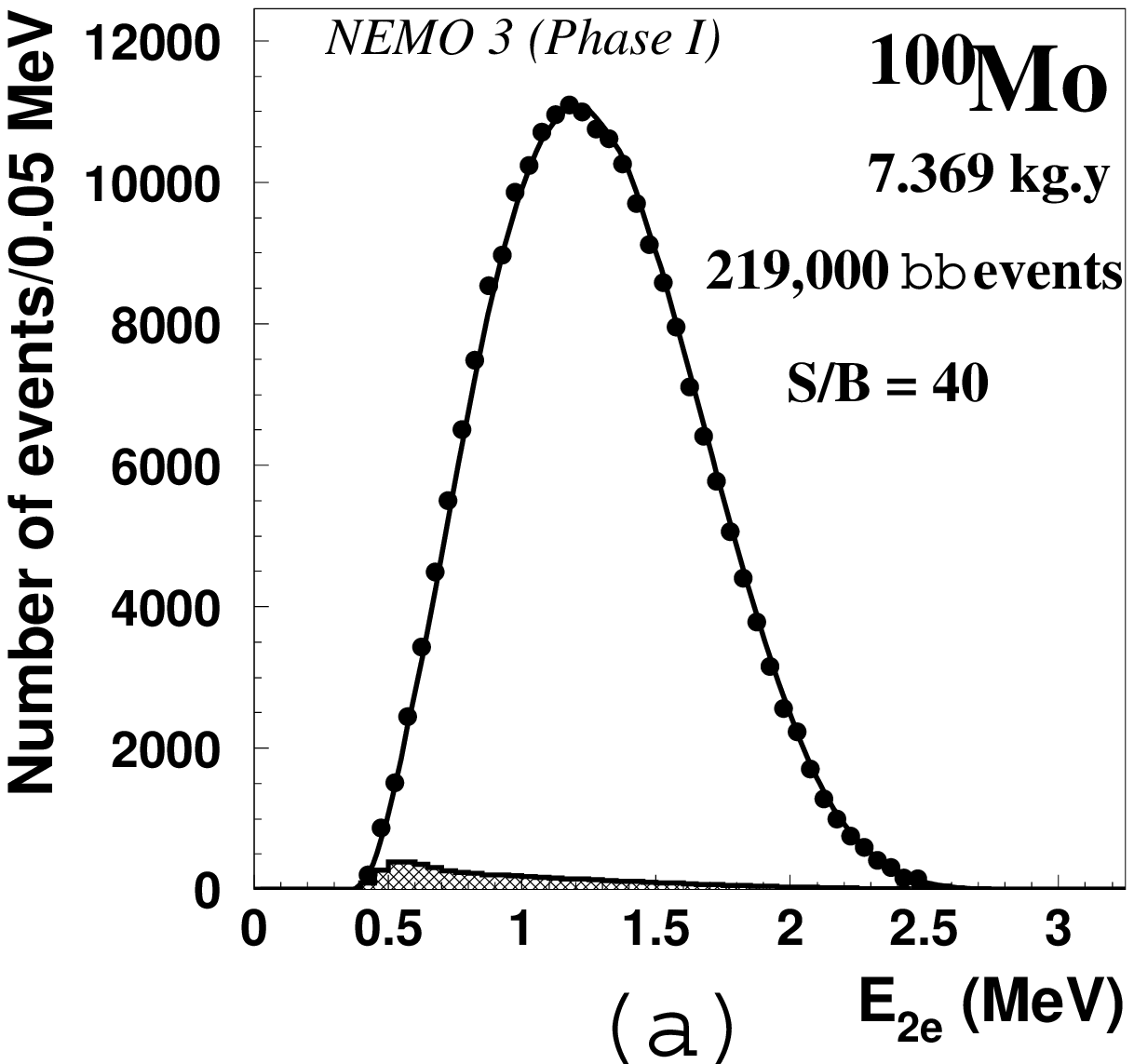}
\includegraphics[scale=0.35]{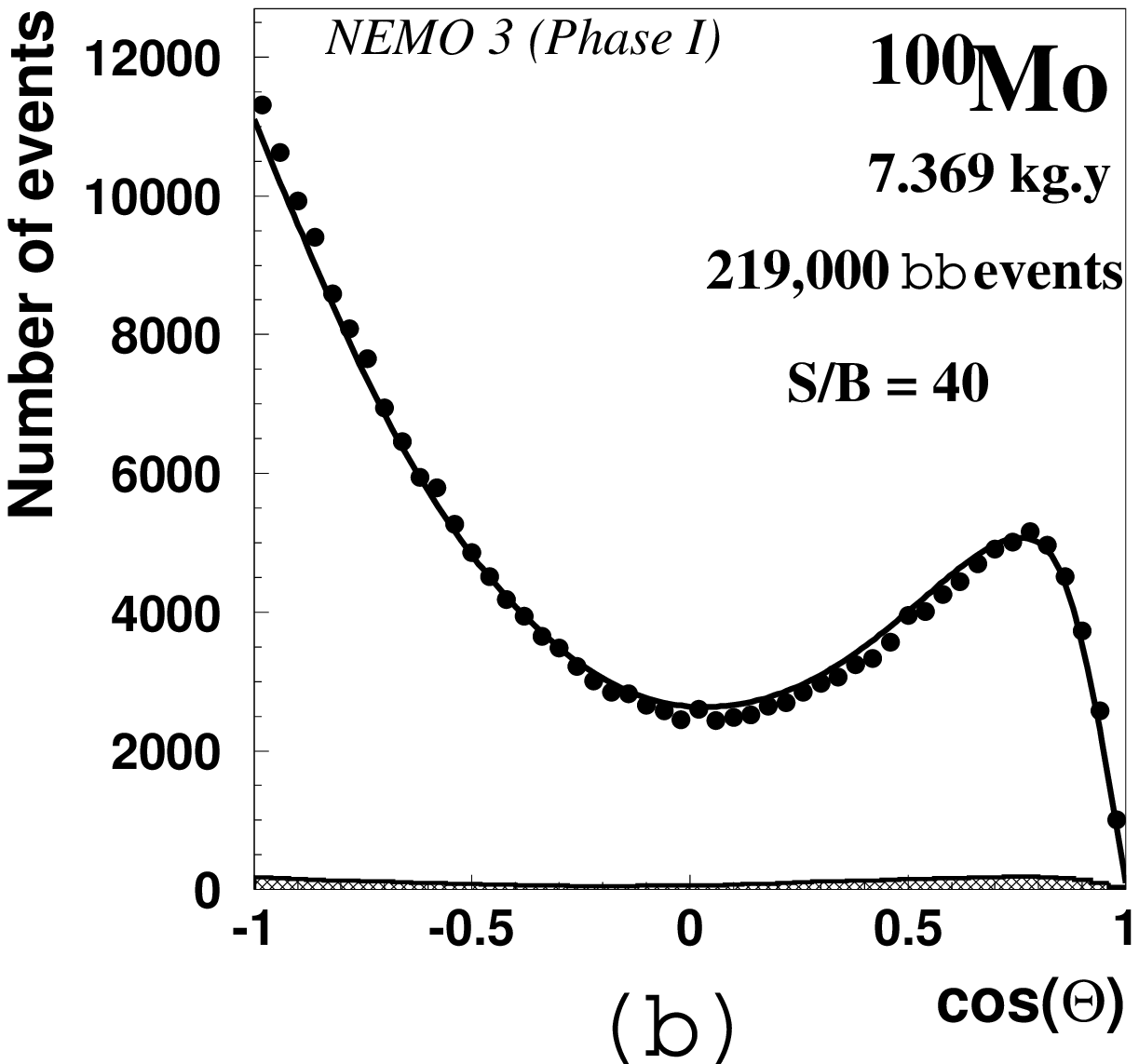}
\includegraphics[scale=0.35]{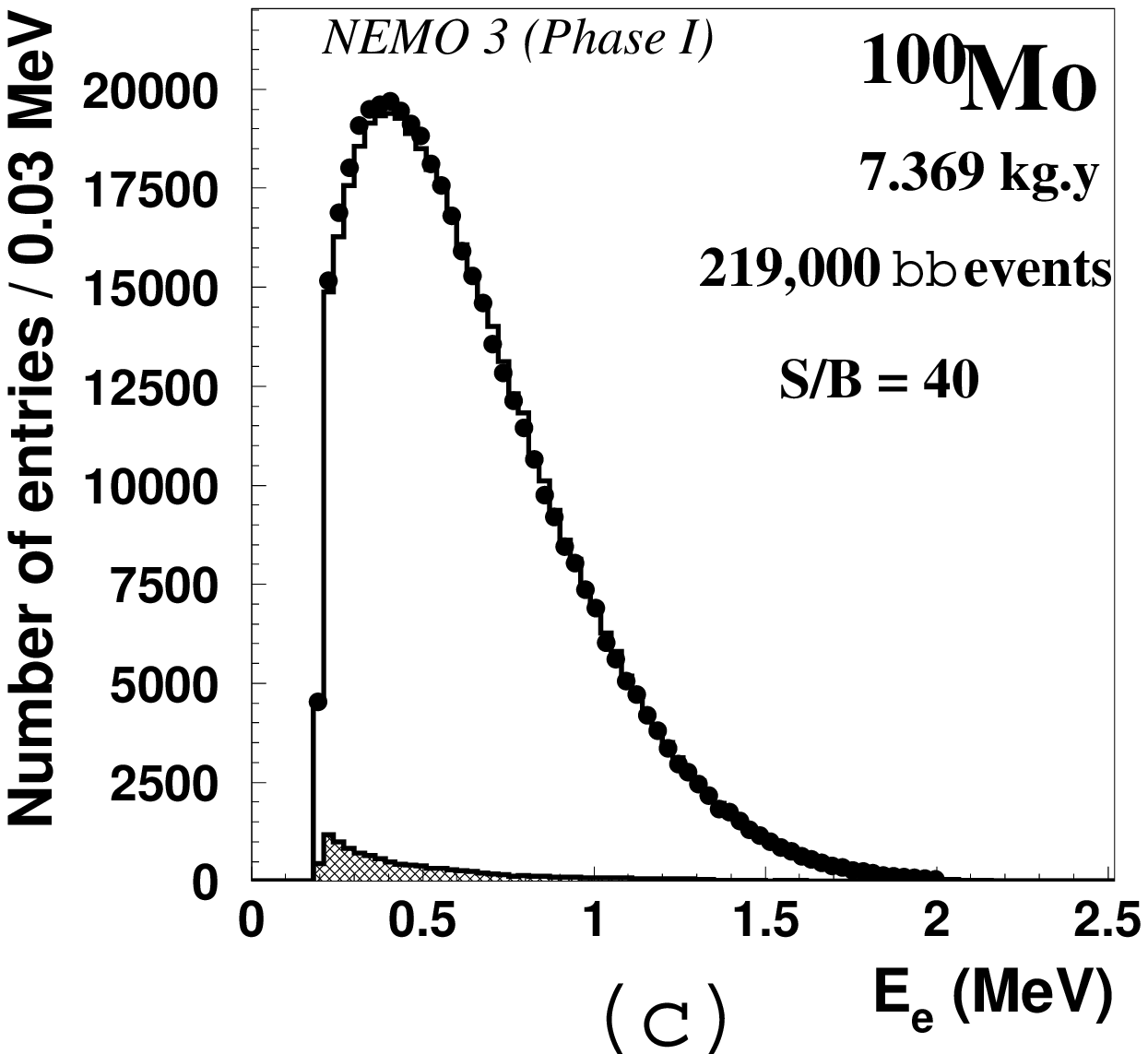}
\end{center}
\caption{(a) Energy sum spectrum of the two
electrons, (b) angular
distribution of the two electrons and (c) single energy spectrum
of the electrons, after background
subtraction from $^{100}$Mo with of 7.369 kg$\cdot$years  
exposure (Phase I) \cite{ARN05a}.
The solid line corresponds to the expected
spectrum from $2\nu\beta\beta$ simulations and the shaded
histogram is the subtracted background
computed by Monte-Carlo simulations.}
\end{figure}

The total number of events exceeds 219,000 which is much greater
than the total statistics of all of the preceding
experiments with $^{100}$Mo (and even greater than the total statistics of all previous 
$2\nu\beta\beta$ decay experiments!). It should also be noted that the
background is as low as $2.5\%$ of the total number of
$2\nu\beta\beta$ events.
The half-life obtained for Phase I data (389 days) is
$T_{1/2}^{2\nu} = [7.11 \pm 0.02(stat) \pm 0.54(syst)] \cdot 10^{18}$ y  
and the single electron energy spectrum is in favour of the Single State Dominance (SSD) 
mechanism (see \cite{SHI06}). The $T_{1/2}$ value is in agreement with previous 
measurements \cite{BAR06} but it has higher precision.

The $2\nu\beta\beta$ decay of $^{100}$Mo to the excited $0^+$ state of $^{100}$Ru (1130 keV) has 
also been measured, $T_{1/2}^{2\nu} = [5.7^{+1.3}_{-0.9}(stat) \pm 0.8(syst)] \cdot 10^{20}$ y \cite{ARN07}.    
The result is in good agreement with previous measurements \cite{BAR95,BAR99,DEB01}.

Fig. 3 (left) shows the tail of the two-electron energy sum spectrum in
the $0\nu\beta\beta$ energy window
for $^{100}$Mo (Phase I+II; 693 days of measurement).
One can see that the experimental spectrum is in good agreement with
the calculated spectrum, which was obtained taking
into account all sources of background. Using a maximum
likelihood method, the following
limits on neutrinoless double beta decay of $^{100}$Mo (mass mechanism; 90\% C.L.) has been obtained:

\begin{equation}
T_{1/2}(^{100}{\rm Mo};0\nu) > 5.8\cdot 10^{23}\; {\rm y}\
\end{equation}

\begin{figure}[htb]
\begin{center}
\includegraphics[width=0.41\textwidth,height=5.5cm]{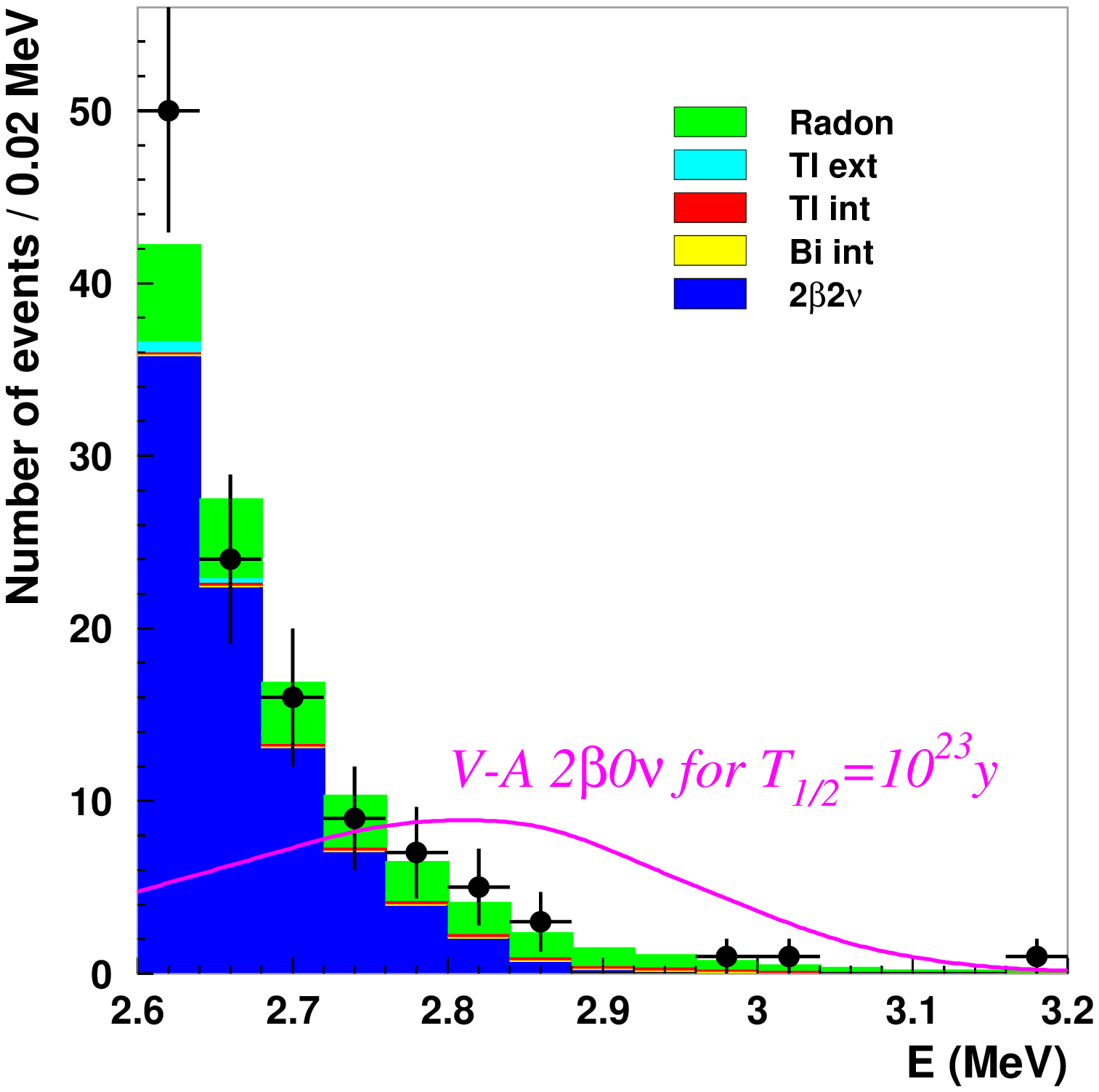}
\includegraphics[width=0.41\textwidth,height=5.5cm]{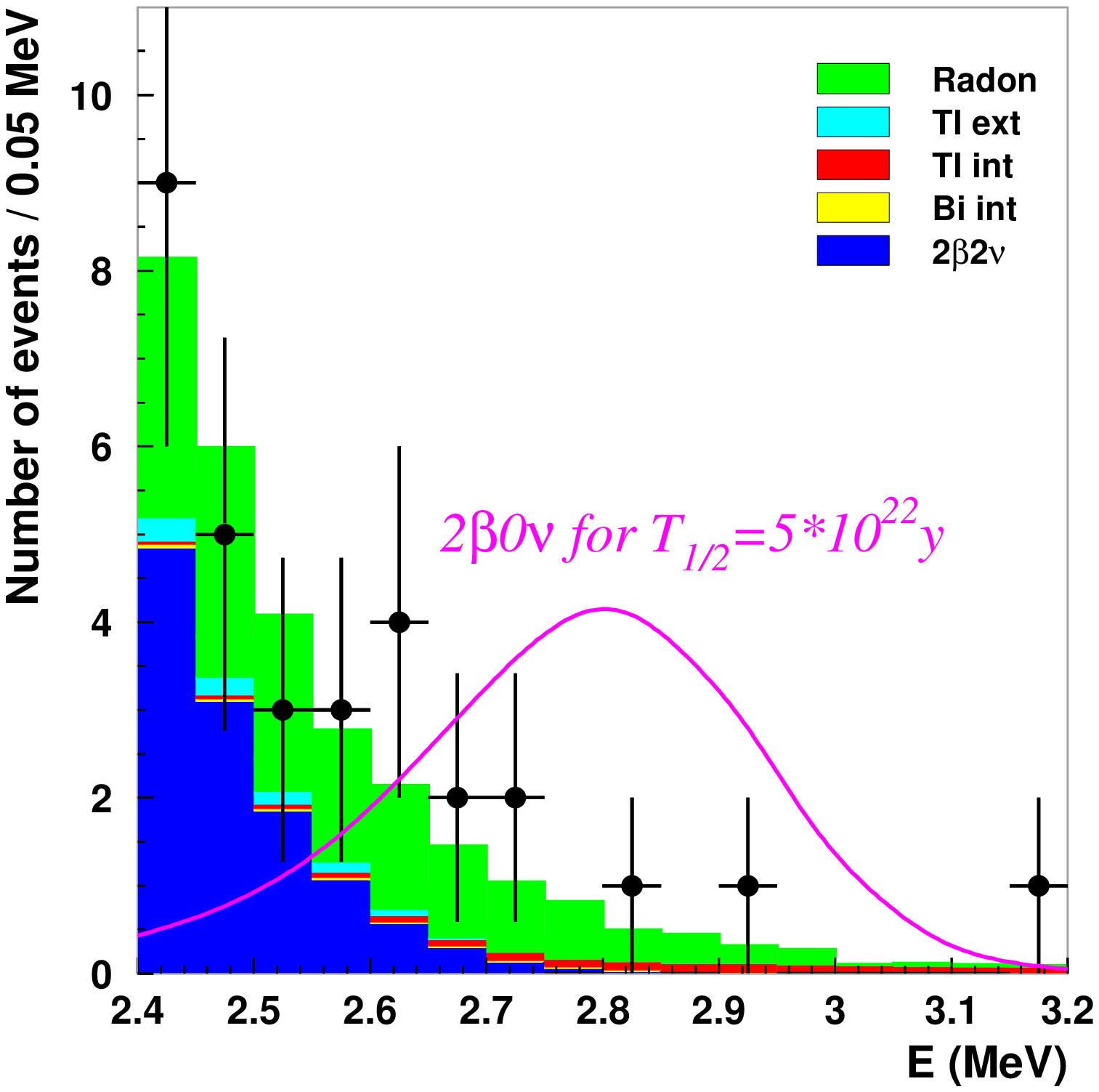}
\end{center}
\caption[Tl]{Distribution of the energy sum of two electrons 
for $^{100}$Mo (left) and  $^{82}$Se (right) (Phase I+II data, 693 days of measurement).}
\label{fig:0nu}
\end{figure}

Using NME values from \cite{KOR07a,SIM08} the bound on
$\langle m_{\nu} \rangle$ gives 0.61-1.26 eV for $^{100}$Mo. 
In this experiment the best present limits on all possible modes
of double beta decay with Majoron emission
have also been obtained \cite{ARN06}.

\subsection{Results for $^{82}$Se ($Q_{2\beta}$ = 2.995 MeV)}

Fig. 4 displays the spectrum of $2\nu\beta\beta$ events
for $^{82}$Se that were collected over 389 days (Phase I) \cite{ARN05a}. 
The signal contains
2,750 2$\nu\beta\beta$ events and the signal-to-background ratio is
4. The measured half-life value is
$T_{1/2}^{2\nu} = [9.6 \pm 0.3(stat) \pm 1.0(syst)] \cdot 10^{19}$ y  
and is also in good agreement with previous measurements \cite{BAR06}. 

\begin{figure}
\begin{center}
\includegraphics[scale=0.6]{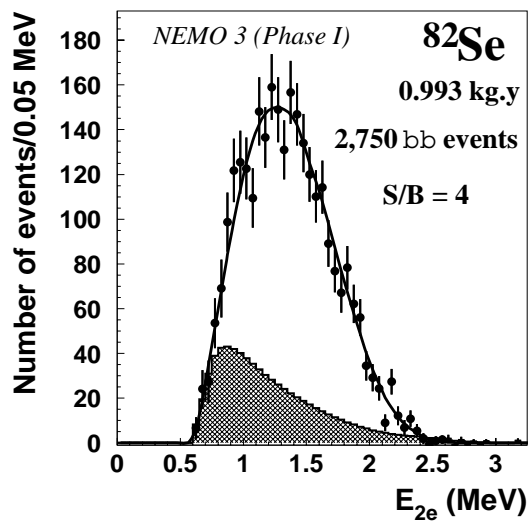}
\end{center}
\caption{Energy sum spectrum of the two
electrons after background subtraction
from $^{82}$Se with 0.993 kg$\cdot$years exposure (same legend as
Fig. 2) \cite{ARN05a}.}
\end{figure}

Fig. 3 (right) shows the tail of the two-electron energy sum spectrum in
the $0\nu\beta\beta$ energy window
for $^{82}$Se (Phase I+II; 693 days of measurement).
One can see that the experimental spectrum is in good agreement with
the calculated spectrum, which was obtained taking
into account all sources of background. Using a maximum
likelihood method, the following
limits on neutrinoless double beta decay of $^{82}$Se (mass mechanism; 90\% C.L.) has been obtained:

\begin{equation}
T_{1/2}(^{82}{\rm Se};0\nu) > 2.1\cdot 10^{23}\; {\rm y}\
\end{equation}

In addition very strong limits on different types of decay with Majoron emission were obtained 
\cite{ARN06}.

\subsection{Results for $^{130}$Te ($Q_{2\beta}$ = 2.529 MeV)}

\begin{figure}
\begin{center}
\includegraphics[width=0.45\textwidth,height=0.45\textwidth,angle=0]{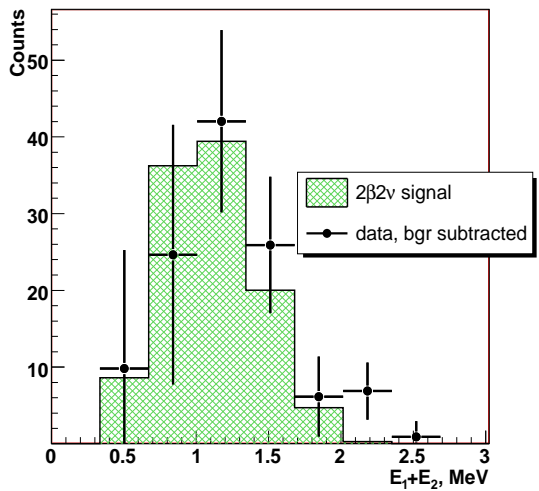}
\end{center}
\caption{\label{fgte} The energy sum spectrum of $^{130}$Te $\beta\beta_{2\nu}$ events with 
background subtracted, Phase II data (534 days).}
\end{figure}

A preliminary measurement of $^{130}$Te half-life is reported here. Due to the extremely
low decay rate, only the low background data (Phase II) was used. In total, 534 days of data
were processed. The number of events observed from the Te source foil is 607, while the
predicted background is 492 two-electron events. A binned maximum likelihood method was used to analyze
the data. A $2\nu\beta\beta$ signal equal
to $109 \pm 21.5 (\rm stat)$ events was found, Fig.~\ref{fgte}. This corresponds to a
 half-life of
$T_{1/2}^{2\nu}(^{130}\rm Te) = [7.6 \pm 1.5(stat) \pm 0.8(syst)] \cdot 10^{20}$~y.

This result is an improvement over the previous attempt at the direct measurement of
$^{130}$Te decay \cite{ARNA03}. It is also in agreement
with predictions based on the geochemical $^{82}$Se/$^{130}$Te ratio and the present
$^{82}$Se decay rate from counting experiments \cite{BAR06}.

Thus it is of particular interest, because there is a disagreement between 
geochemical measurements. One group found $T_{1/2} \approx 8 \cdot 10^{20}$~y
\cite{geo_young}, while the other gives $T_{1/2}\approx (2.5-2.7) \cdot 10^{21}$~y \cite{geo_old}. Also
it was noted that a smaller $T_{1/2}$ value was obtained in the experiments with "young"
ores (younger than 100 million years). This led to the hypothesis that the differences can
be accounted for by
variations of the Fermi constant $\rm G_F$ with time
\cite{Gvariation}. The NEMO 3 measurement
reported does not contradict this hypothesis, and suggests that the possible $\rm G_F$
variation should be tested
using geochemical methods for other $\beta\beta$ decaying nuclei.

\subsection{Results for $^{150}$Nd ($Q_{2\beta}$ = 3.367 MeV)}

A preliminary measurement of the half-life of $^{150}$Nd was obtained for a 46.6 g of 
$^{150}$Nd$_2$O$_3$ (enrichment of $(91 \pm 0.5)\%$ and the weight of $^{150}$Nd is 36.6 g) from data 
collected between February 2003 and December 2006 corresponding to 939 days of data during 
the Phases I and II of the experiment. The 2828 $\beta\beta$ type events were observed with a 
signal-over-background ratio of 2.8. The distribution of the energy sum of the electrons in 
$2\nu\beta\beta$ type events and their angular distribution are shown in Fig. 6. The background 
subtracted data and the $2\nu\beta\beta$ signal expectation obtained from Monte Carlo calculations 
are in good agreement. The $2\nu\beta\beta$ selection efficiency is 7.2\%. The measured 
half-life is $T_{1/2}^{2\nu}(^{150}\rm Nd) = [9.20 ^{+0.25}_{-0.22}(stat) \pm 0.62(syst)] \cdot 10^{18}$~y. 
This value is between the two previous results obtained from experiments with time projected 
chambers, $T_{1/2}^{2\nu} = [18.8 ^{+6.6}_{-3.9}(stat) \pm 1.9(syst)] \cdot 10^{18}$~y 
\cite{ART95} and $T_{1/2}^{2\nu} = [6.75 ^{+0.37}_{-0.42}(stat) \pm 0.68(syst)] \cdot 10^{18}$~y 
\cite{MOE97}, but has much better statistical error.

\begin{figure}
\begin{center}
\includegraphics[scale=0.3]{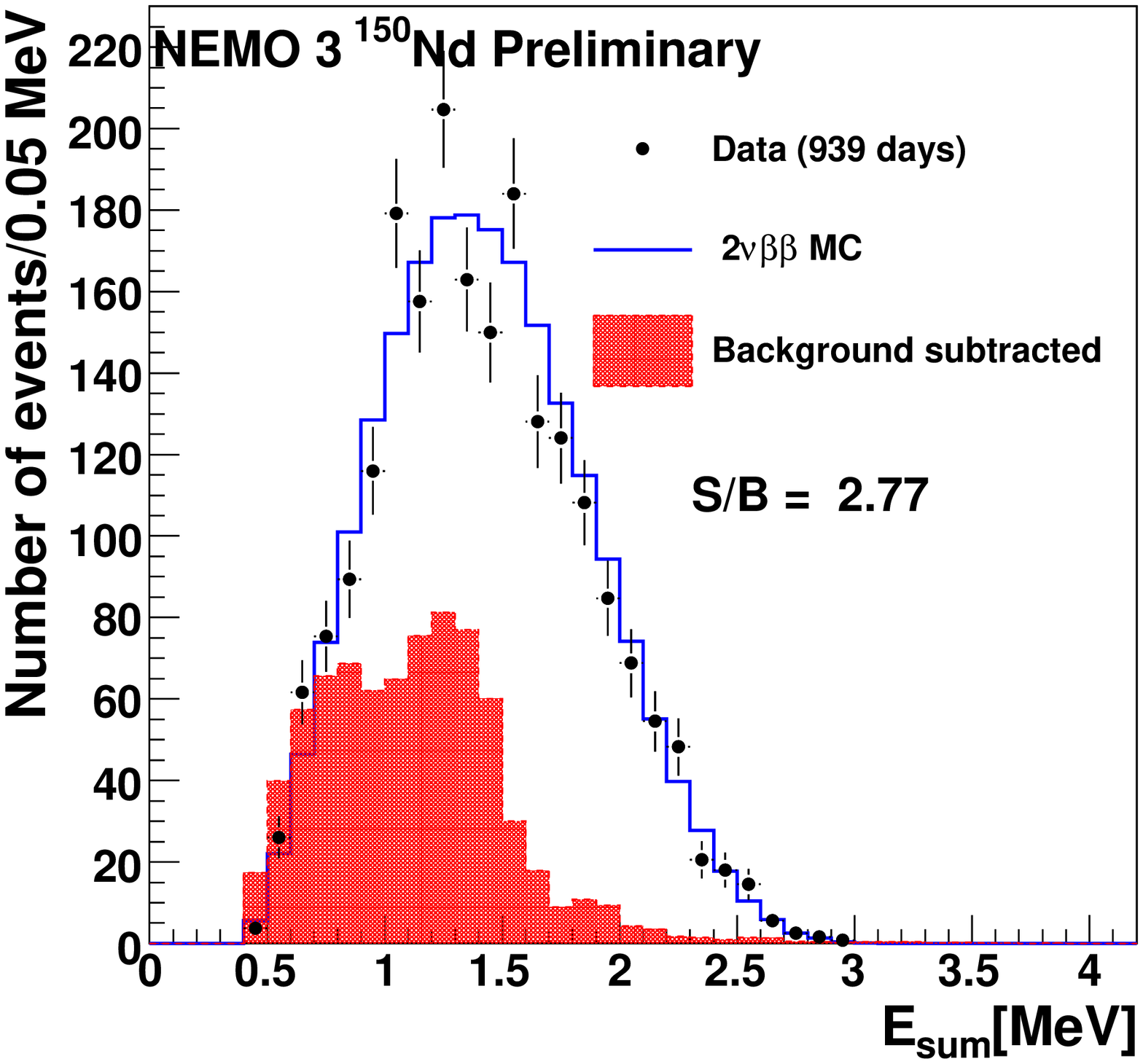}
\includegraphics[scale=0.3]{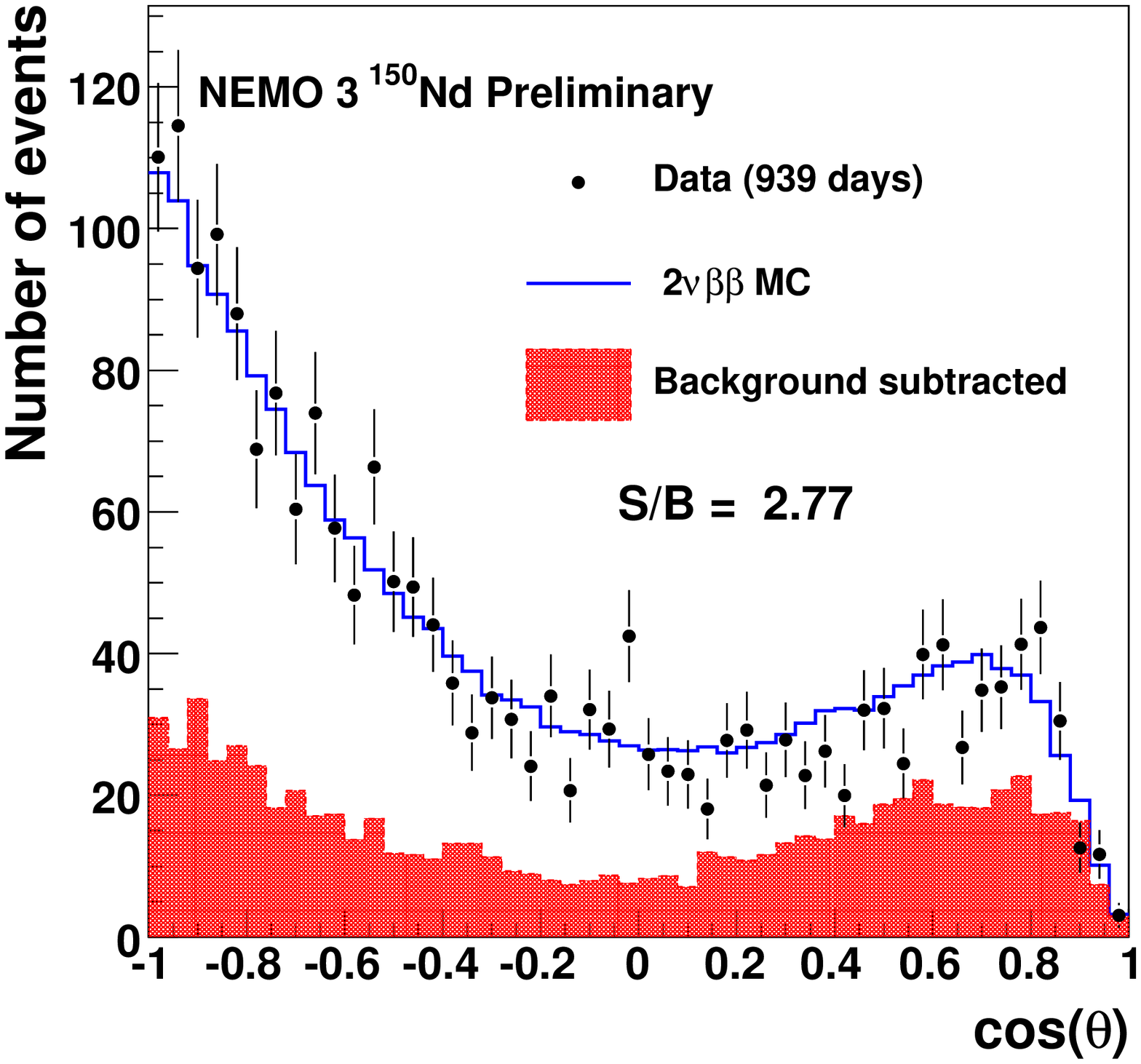}
\end{center}
\caption{The energy sum spectrum (left) and 
the angular distribution (right) of $^{150}$Nd $2\nu\beta\beta$ events with 
background subtracted Phase I+II data (939 days).}

\end{figure}

A $0\nu\beta\beta$ decay signal would correspond to an excess of $\beta\beta$ type events in 
the distribution of the energy sum of the electrons around the energy of the transition $Q_{\beta\beta}$. No excess of events in the distribution of the energy 
sum of the electrons from $\beta\beta$ type events originating from the $^{150}$Nd sample was observed during the 
939 days of data collection. A limit on the half-life of the $0\nu\beta\beta$ process is subsequently set
using the $CL_s$ method \cite{JUN99} for $E_{sum} > 2.5$ Mev. The limit on the half-life is $T_{1/2}(0\nu) > 1.8\cdot 10^{22}$ y 
at a 90\% C.L., which translates into an upper limit $\langle m_{\nu} \rangle < 1.7-2.4$ eV using the 
NME from QRPA calculations (deformation is not taken into account) \cite{ROD07} or $\langle m_{\nu} \rangle < 4.8-7.6$ eV using 
the NME from pseudo-SU(3) model (developed for deformed nuclei) \cite{HIR95}. The limit on the half-life was improved 
when compared 
to the previous result, $T_{1/2}(0\nu) > 1.7\cdot 10^{21}$ y (95\% C.L.) \cite{KLI86}. 

In the assumption of a $0\nu\beta\beta$ process involving right currents (V+A), the limit is found to be
$T_{1/2}(0\nu) > 1.27\cdot 10^{22}$ y at a 90\% C.L. For a $0\nu\beta\beta$ process with Majoron emission 
(spectral index n = 1) the obtained limit is $T_{1/2}(0\nu\chi^0) > 1.55\cdot 10^{21}$ y (90\% C.L.).

\subsection{Results for $^{48}$Ca ($Q_{2\beta}$ = 4.272 MeV)}

A preliminary measurement of the half-life of $^{48}$Ca was obtained for 17.5 g of $^{48}$CaF$_2$ 
(enrichment $(73.2 \pm 1.6)\%$ and the weight of $^{48}$Ca is 7 g) from the data 
corresponding to 943.16 days of data collection during the Phases I and II.

\begin{figure}
\includegraphics[width=0.8\textwidth,height=0.6\textwidth,angle=0]{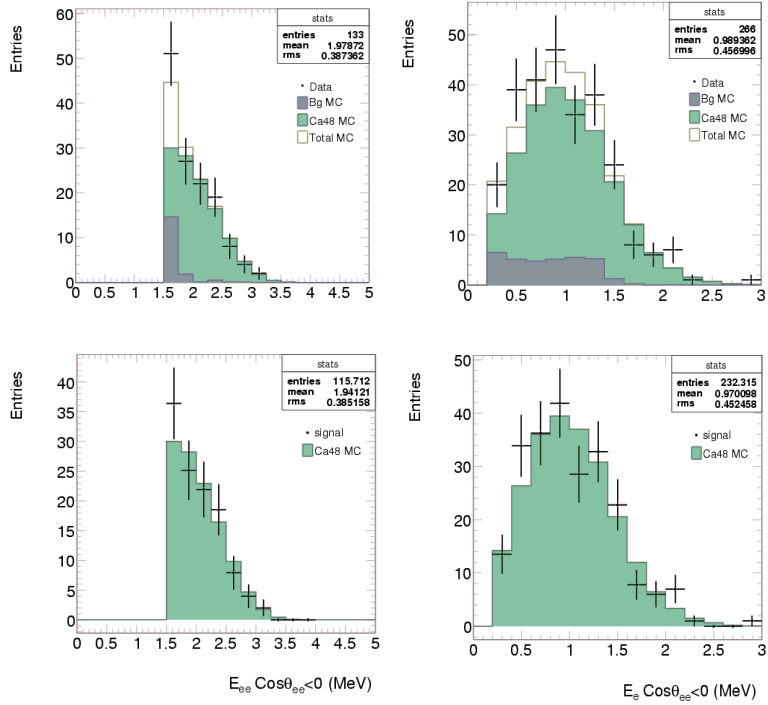}
\caption{Spectra obtained with $^{48}$Ca (Phase I+II; 943.16 days). Energy sum spectrum of the two
electrons without background subtraction (top-left); single energy spectrum of the two
electrons without background subtraction (top-right); energy sum spectrum of the two
electrons after background subtraction (bottom-left); single energy spectrum of the two
electrons after background subtraction (bottom-right).}
\end{figure}

The sample of $^{48}$Ca inside NEMO~3 is known to be contaminated with $^{90}$Sr, which is the main background source
in this case. The activity for the $^{90}$Sr contamination as measured by the NEMO~3 detector is 1699 $\pm$ 3(stat) mBq/kg.
To suppress the $^{90}$Sr background contribution $\beta\beta$ events with a sum energy $> 1.5$ MeV and cos$\theta < 0$ 
were selected. Fig. 7 shows the two electron sum spectrum and single electron spectrum after these cuts. 
After 22636 h of data collection, a total of 133 data events were selected, and background was estimated as 17.13 events. 
The efficiency from MC is 3.3\%. The obtained half-life value is, 
$T_{1/2}^{2\nu} = [4.4 ^{+0.5}_{-0.4}(stat) \pm 0.4(syst)] \cdot 10^{19}$~y. This value is in good agreement with two 
previous measurements \cite{BAL96,BRU00}, and has much higher precision. The half-life value obtained is in good agreement 
with the Shell Model prediction of $T_{1/2}^{2\nu} = 3.9 \cdot 10^{19}$~y \cite{CAU99}.

In the search for the $0\nu\beta\beta$ decay of $^{48}$Ca, the $CL_s$ method \cite{JUN99} was employed. The range 
$E_{ee} > 2.0$ MeV was used for the limit calculation. The limit obtained is $T_{1/2}(0\nu) > 1.3\cdot 10^{22}$ y (90\% C.L.).

Using the NME from \cite{CAU08}, the corresponding limit on the effective Majorana neutrino mass is 
$\langle m_{\nu} \rangle < 29.7$ eV. The limit obtained is approximately the same as the best previous result, 
$T_{1/2}(0\nu) > 1.4\cdot 10^{22}$ y (90\% C.L.) \cite{OGA04}.

\subsection{Results for $^{96}$Zr ($Q_{2\beta}$ = 3.350 MeV)}

A preliminary measurement of the half-life of $^{96}$Zr was obtained for a 22 g sample of $^{96}$ZrO$_2$ 
(enrichment $(57.3 \pm 1.4)\%$ and the weight $^{96}$Zr is 9.4 g) from the data 
corresponding to 924.67 days of data collection during the Phases I and II.
Fig. 8 shows the two electron sum spectrum, single electron spectrum and angular distribution after subtraction of background. 
These were 331  $2\nu\beta\beta$ events selected (best fit). 
The efficiency from MC is 7.6\%. The  half-life obtained is, 
$T_{1/2}^{2\nu} = [2.3 \pm 0.2(stat) \pm 0.3(syst)] \cdot 10^{19}$~y. This value is in good agreement with  
previous NEMO~2 measurement \cite{ARN99}, but again has much higher precision.

\begin{figure}
\begin{center}
\includegraphics[scale=0.3]{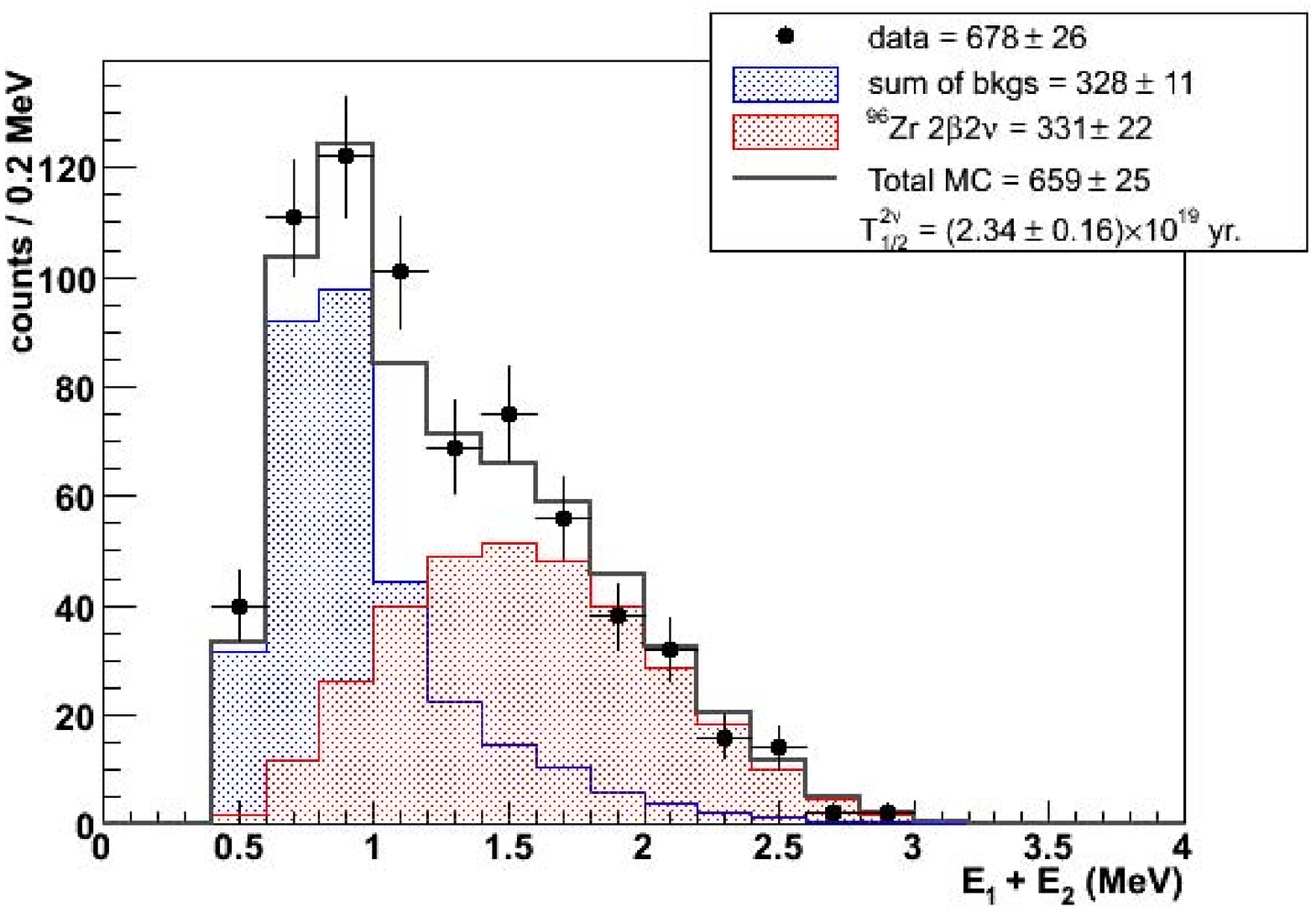}
\includegraphics[scale=0.3]{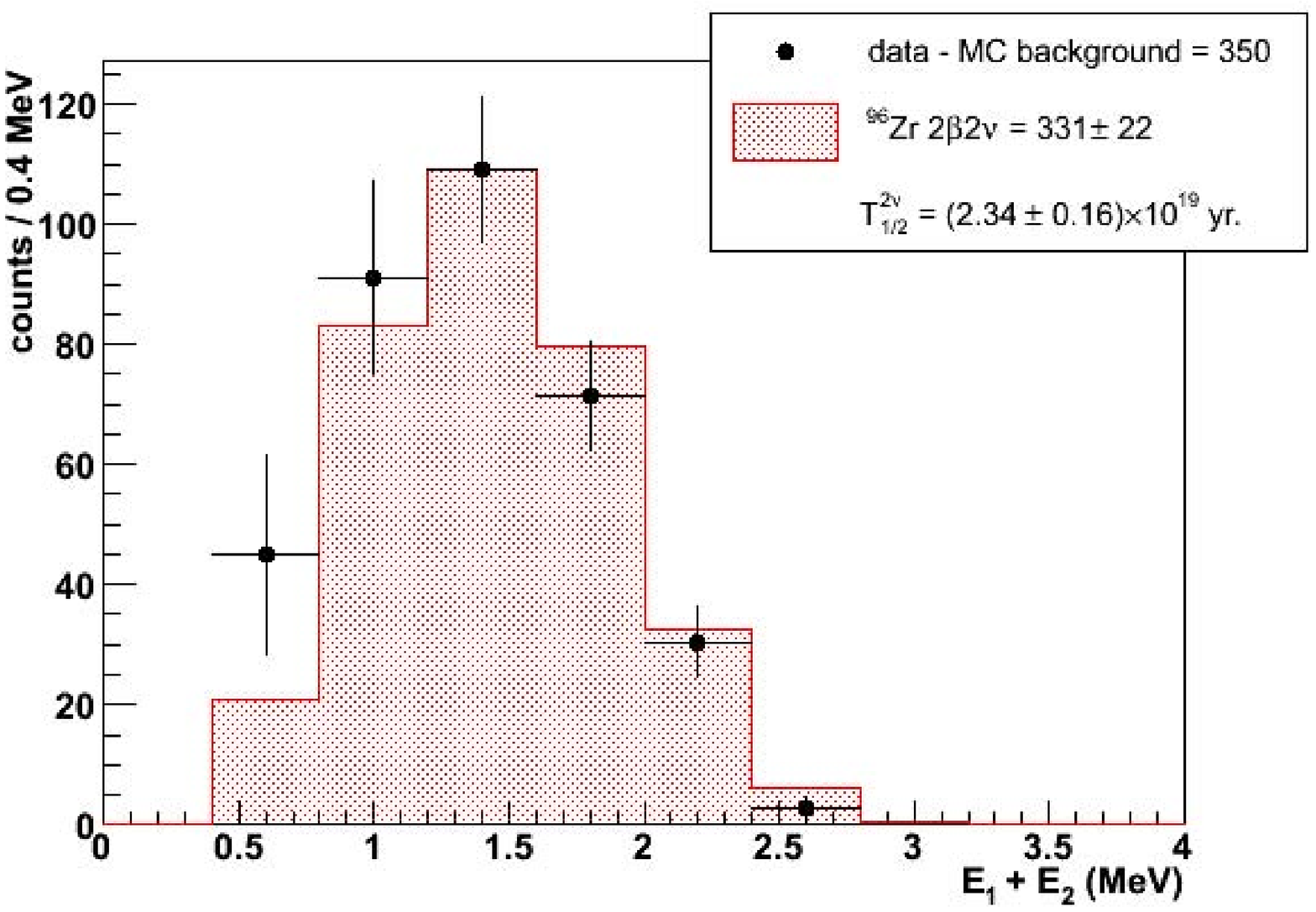}\\
\includegraphics[scale=0.3]{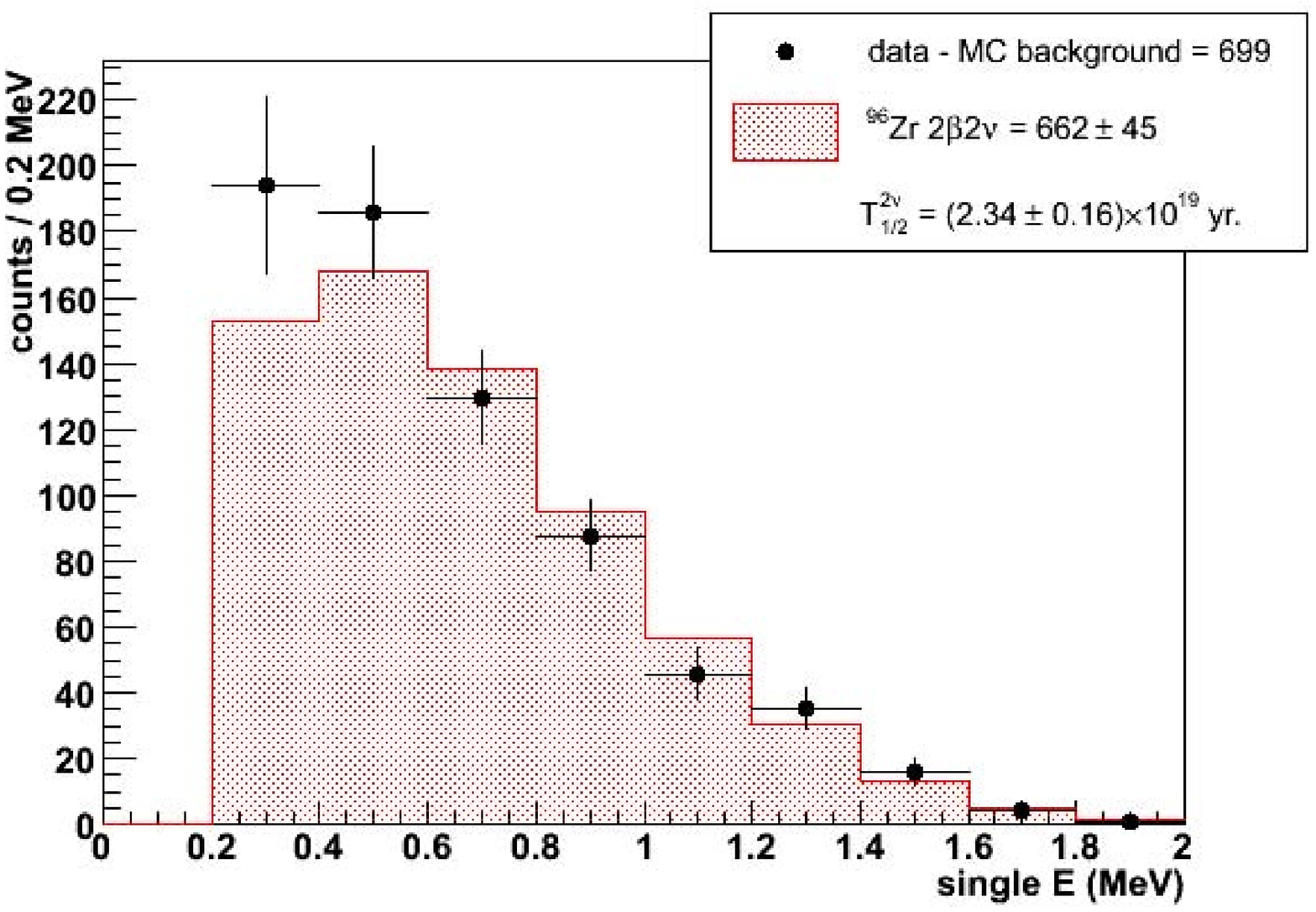}
\includegraphics[scale=0.3]{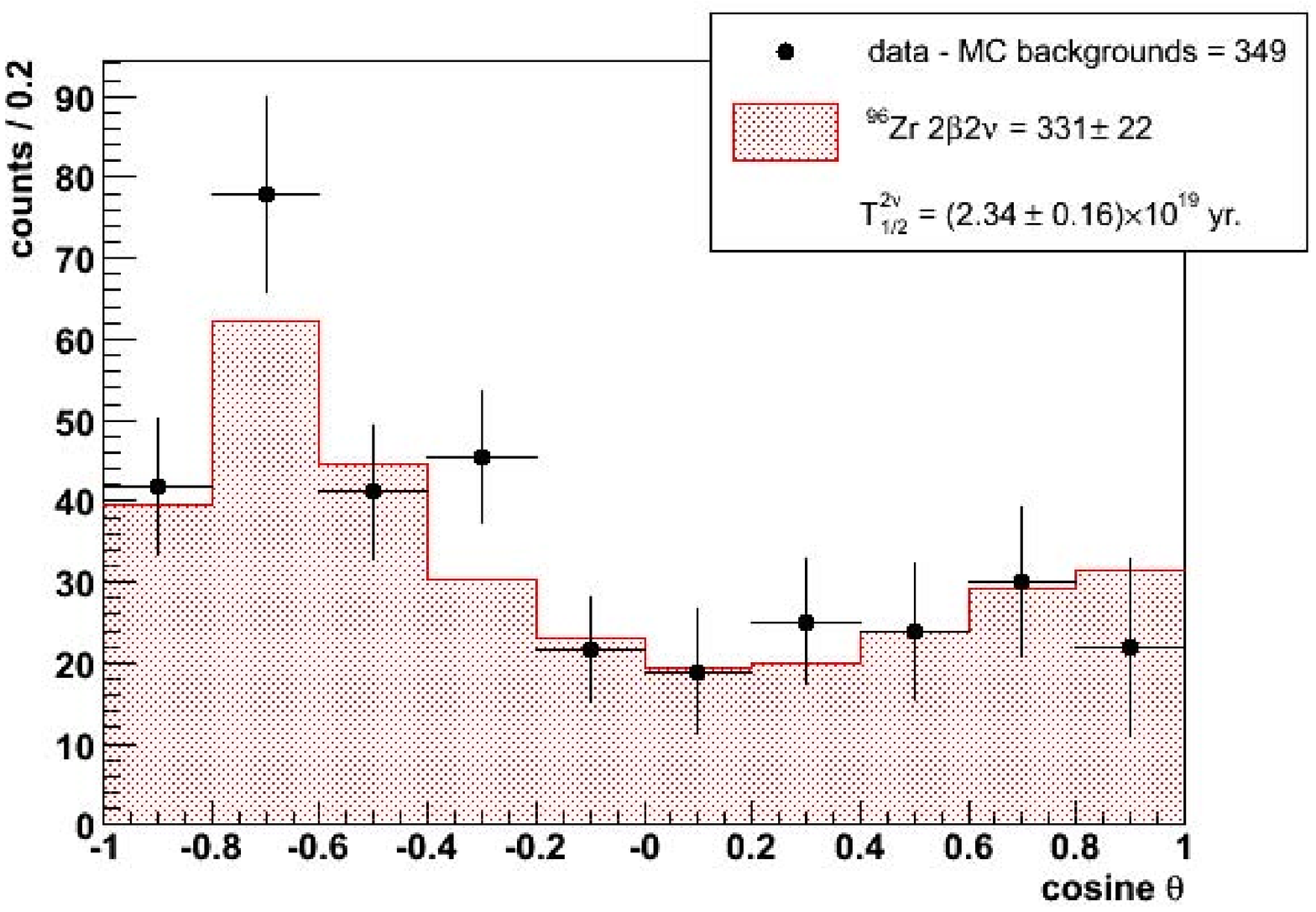}
\end{center}
\caption{Spectra obtained with $^{96}$Zr (Phase I+II; 924.67 days). The energy sum spectrum of the two
electrons without background subtraction (top-left); energy sum spectrum of the two
electrons after background subtraction (top-right); single energy spectrum
of the electrons, after background subtraction (bottom-left); angular
distribution of the two electrons after background subtraction (bottom-right)}
\end{figure}

For neutrinoless decay (mass mechanism) the limit is $T_{1/2}(0\nu) > 8.6\cdot 10^{21}$ y (90\% C.L.) 
and the corresponding limit on the effective Majorana neutrino mass is $\langle m_{\nu} \rangle < 7.4-20.1$ eV (using 
NME values from \cite{KOR07a,SIM08}). This limit on the half-life is in 8.6 times better than the previous 
NEMO~2 result \cite{ARN99}.

\begin{table}[ht]
\label{Table6}
\caption{Two neutrino half-life values for different nuclei
obtained in
the NEMO-3 experiment.  
S/B is the signal-to-background ratio.}
\vspace{0.5cm}
\begin{center}
\begin{tabular}{ccccc}
\hline
Isotope & Measurement & Number of & S/B & $T_{1/2}(2\nu)$, y \\
& time, days & $2\nu$ events & &  \\
\hline
$^{100}$Mo & 389 & 219000 & 40 & $[7.11 \pm 0.02(stat) \pm
0.54(syst)]\cdot 10^{18}$ \cite{ARN05a}  \\
$^{100}$Mo- & 334.3 & 37.5 & 3 & $[5.70^{+1.3}_{-0.9}(stat) \pm 0.8(syst)]
\cdot 10^{20}$ \cite{ARN07}  \\ 
$^{100}$Ru(0$^+_1$) & & & & \\
$^{82}$Se & 389 & 2750 & 4 & $[9.6 \pm 0.3(stat) \pm 1.0(syst)]
\cdot 10^{19}$ \cite{ARN05a} \\
$^{116}$Cd & 168.4 & 1371 & 7.5 & $[2.8 \pm 0.1(stat) \pm
0.3(syst)]\cdot 10^{19}$  \\
$^{48}$Ca & 943.16 & 116 & 6.8 & $[4.4^{+0.5}_{-0.4}(stat) \pm 0.4(syst)]
\cdot 10^{19}$  \\
$^{96}$Zr & 924.67 & 331 & 1 & $[2.3 \pm 0.2(stat) \pm 0.3(syst)]
\cdot 10^{19}$  \\
$^{130}$Te & 534 & 109 & 0.2 & $[7.6 \pm 1.5(stat) \pm 0.8(syst)]
\cdot 10^{20}$  \\
$^{150}$Nd & 939 & 2018 & 2.8 & $[9.20^{+0.25}_{-0.22}(stat) \pm 0.62(syst)]
\cdot 10^{18}$  \\
\hline
\end{tabular}
\end{center}
\end{table}

\begin{table}[ht]
\label{Table6}
\caption{Limits at 90\% C.L. on $0\nu\beta\beta$ decay (neutrino mass mechanism) for different nuclei
obtained in
the NEMO-3 experiment.}
\vspace{0.5cm}
\begin{center}
\begin{tabular}{ccc}
\hline
Isotope & Measurement & $T_{1/2}(0\nu)$, y \\
& time, days &  \\
\hline
$^{100}$Mo & 693 & $ > 5.8\cdot 10^{23}$ y  \\
$^{82}$Se & 693 & $ > 2.1\cdot 10^{23}$ y  \\
$^{116}$Cd & 77 & $> 1.6\cdot 10^{22}$ y   \\
$^{48}$Ca & 943.16 & $> 1.3\cdot 10^{22}$ y   \\
$^{96}$Zr & 924.67 & $> 8.6\cdot 10^{21}$ y  \\
$^{150}$Nd & 939 & $> 1.8\cdot 10^{22}$ y   \\
\hline
\end{tabular}
\end{center}
\end{table}

\section{Conclusion}

The NEMO~3 detector has been operating within the target performance specifications since February 2003. 
The $2\nu\beta\beta$ decay  has been measured for a seven isotopes with high statistics and greater 
precision than previously. The $^{100}$Mo $2\nu\beta\beta$ decay to the $0^+_1$ excited state of $^{100}$Ru 
has also been measured. All obtained results are presented in table 2. First three results are published and others
are preliminary. 
No evidence for $0\nu\beta\beta$ decay was found for all seven isotopes. Obtained limits are presented in table 3.
The best limits have been obtained with $^{100}$Mo ($ > 5.8\cdot 10^{23}$ y at 90\% C.L.) 
and $^{82}$Se ($ > 2.1\cdot 10^{23}$ y at 90\% C.L.). After five years of data collection the expected 
sensitivity at the 90\% C.L. will be $ \sim 2\cdot 10^{24}$ y for $^{100}$Mo, $ \sim 8\cdot 10^{23}$ y for $^{82}$Se 
and $\sim 10^{22}-10^{23}$ y for the other isotopes.

\section{Acknowledgments}

This work was supported by Russian Federal Agency for Atomic Energy and by Federal Agency for Science and Innovations
(contract No. 02.516.11.6099). 
A portion of this work was supported by grants from RFBR (No. 06-02-72553).

\end{document}